\documentclass[11pt,a4paper]{article}
\usepackage[hyperref]{acl2021}
\usepackage{times}
\usepackage{latexsym}

\usepackage{graphicx}

\usepackage{microtype}

\aclfinalcopy 

\title{Malware Knowledge Graph Generation}

\author{
Sharmishtha Dutta, Nidhi Rastogi*, Destin Yee, Chuqiao Gu, Qicheng Ma\\
\texttt{duttas,raston2,yeed3,guc,maq3@rpi.edu}\\
Rensselaer Polytechnic Institute, Troy, NY
}

\begin{document}
\maketitle
\begin{abstract}
Cyber threat and attack intelligence information are available in non-standard format from heterogeneous sources. Comprehending them and utilizing them for threat intelligence extraction requires engaging security experts. Knowledge graphs enable converting this unstructured information from heterogeneous sources into a structured representation of data and factual knowledge for several downstream tasks such as predicting missing information and future threat trends. Existing large-scale knowledge graphs mainly focus on general classes of entities and relationships between them. Open-source knowledge graphs for the security domain do not exist. To fill this gap, we've built \textsf{TINKER} - a knowledge graph for threat intelligence (\textbf{T}hreat \textbf{IN}telligence \textbf{K}nowl\textbf{E}dge g\textbf{R}aph). \textsf{TINKER} is generated using RDF triples describing entities and relations from tokenized unstructured natural language text from 83 threat reports published between 2006-2021. We built \textsf{TINKER} using classes and properties defined by open-source malware ontology and using hand-annotated RDF triples. We also discuss ongoing research and challenges faced while creating \textsf{TINKER}.
\end{abstract}

\section{Introduction}
There is no dearth of cybersecurity threat and attack information on the Internet. Both structured and unstructured data is available to security professionals and researchers to help in their increasingly complex jobs of finding zero-day attacks, protecting intellectual property, and preventing intrusions from adversaries. CVE~\cite{martin2007common}, NVD~\cite{booth2013national} are vulnerability tracking programs where the security community feeds information. For sharing threat intelligence, there are industry standards like Structured Threat Information eXpression (STIX)~\cite{stix}, which provides a language-agnostic framework to capture threat intelligence into a shareable package. In contrast, the trusted automated exchange of indicator information (TAXII)~\cite{connolly2014trusted} is a platform that can send and receive the STIX package. Here, despite concerted efforts by several to organize malware threat information over the past decade, we face two challenges that need immediate attention. The first is the varying degrees of data shared amongst organizations using the same standard. The other challenge, and the one we will address in this paper, is the lack of information standardization and data contextualization.
\par
Security ontologies (specifically malware) have attempted to address the first concern \cite{uco}, \cite{swimmer} primarily to collate data in a structured format and distribute it \cite{stix,connolly2014trusted}. Most existing approaches have a specific purpose of identifying attack patterns, vulnerability, threat, or information dissemination. This paper builds on the popular and yet limited strengths of current security standards, domain linkage properties of the semantic web, and natural language domains. Security ontologies \cite{rastogi2020malont}, \cite{swimmer} are extended to extract cybersecurity threat intelligence from online sources.  Other benefits include interoperability of information across applications and creating labeled features for machine learning models. When data is labeled, using ontologies allows to run queries over large datasets, identify trends, and predict future occurrences.

\par

In this paper, we propose \textsf{TINKER}, a hand-curated knowledge graph that extracts information from unstructured threat-related data. \textsf{TINKER} converts information into a standardized structured format called RDF triples. Triples support the inference, which allows the automated discovery of new facts based on a combination of data and rules for understanding that data. Therefore, \textsf{TINKER} can discover new threat intelligence since it combines malware data using triples. It also connects with general world concepts, thus blending "knowledge," correlating, grouping concepts, and inferring new insights.

\par

We instantiate existing malware ontologies using over 83 threat documents written between 2006-2021 downloaded from the Internet in this work. Approximately 3000 triples were created following malware ontology in \cite{rastogi2020malont,swimmer}  and using the Brat annotation tool\footnote{https://brat.nlplab.org/}. The ontology provided the classes' names, which are called entities in \textsf{TINKER}. The ontology also contains properties called relations in \textsf{TINKER}, which define the relationships between different classes. Security experts on the research team then verified the annotated dataset. Triples generated using these entities are the structural components of \textsf{TINKER} that capture facts related to malware threat intelligence. Aside from that, triples allow reasoning over them, leading to new information missing earlier or was inaccurate. For example, consider the following snippet from a threat report:\footnote{https://www.lastwatchdog.com/wp/wp-content/uploads/Saudi-Arabia-Dustman-report.pdf}

\begin{center}
\textit{``...DUSTMAN can be considered as a new variant of ZeroCleare malware...both
DUSTMAN and ZeroCleare utilized a skeleton of the modified “Turla Driver Loader (TDL)”...The malware executable file “dustman.exe” is not the actual wiper, however, it contains all
the needed resources and drops three other files [assistant.sys, elrawdisk.sys, agent.exe] upon execution...”}
\end{center}

\begin{table}[!t]
\renewcommand{\arraystretch}{1.2}
\caption{Triples extracted from sample threat report text.}
\label{table:triple_list}
\small
\centering
\begin{tabular}{|c| c| c|}
\hline
Head & Relation & Tail\\
\hline
$\langle$DUSTMAN, & similarTo, & ZeroCleare$\rangle$\\
$\langle$DUSTMAN, & involves, & Turla Driver Loader(TDL)$\rangle$\\
$\langle$ZeroCleare, & involves, & Turla Driver Loader(TDL)$\rangle$\\
$\langle$DUSTMAN, & involves, & dustman.exe$\rangle$\\
\hline
\end{tabular}
\end{table}

Table \ref{table:triple_list} shows a set of triples generated from this text. Together, many such triples form the malware knowledge graph where the entities and relations model nodes and directed edges, respectively.

\par
Creating \textsf{TINKER} from scratch was no trivial task, and therefore we discuss the challenges that went in this effort. Nonetheless, what makes the knowledge graph generation a focused and finite effort is malware ontology to build it. A set of competency questions help create the ontology, which the ontology should be able to answer. Threat reports have concepts that are extracted by ontology defined classes. Also included are the potential relationships between concepts (or entities in knowledge graphs) and what we find essential in predicting from the information in hand. When combined with linked open data (LOD)\cite{bauer2011linked}, \textsf{TINKER} improves the semantic content of the data and link datasets at a schema level. It also ensures interoperability with other ontologies and knowledge graphs in the future.

\section{Background Work and Motivation}
Several generic and specific knowledge graphs such as Freebase\cite{bollacker2008freebase}, Google's Knowledge Graph\footnote{https://developers.google.com/knowledge-graph}, and WordNet\cite{fellbaum2010wordnet} are available as open source. Linked Open data\cite{bauer2011linked} refers to all the data available on the Web, accessible using standards such as Uniform Resource Identifiers (URIs) to identify entities and Resource Description Framework (RDF) to unify entities into a data model. However, it does not contain domain-specific entities that deal with cyber threats and intelligence. Other existing knowledge graphs are either not open source, too specific, or do not cater to the breadth of information our research intends to gather from intelligence reports.
\par

As part of entity recognition of key phrases in the threat reports, we explored DBpedia Spotlight \cite{mendes2011dbpedia}. It enables configuration of annotations in domain-specific text, such as cybersecurity) by using a few knowledge bases (as of 2021), including DBpedia, FreeBase (deprecated), and Schema.org, while offering quality measure on disambiguation confidence. Each annotated key phrase also gets an ontology URI that assures provenance to the concepts. However, when annotating text using these ontologies, the key-phrase miner from DBPedia Spotlight~\cite{mendes2011dbpedia} misses out on context determination. For example, a threat report on \textit{"Operation Aurora"} was correctly annotated as a \textsf{Malware} class from the Wikipedia derived ontology. However, the threat report instead uses \textit{"Aurora"} for a majority of the report. We are currently exploring entity extraction approaches that are semantically and syntactically similar. Therefore, a simple CRF or n-gram based approach, despite the high performance, will lead to false-positive annotations. Also, it was missing relation extraction from annotated entities. Contextual entity extraction is exceptionally crucial for cybersecurity threat intelligence due to the continually shifting and evolving landscape. In the language of knowledge graphs, this means \textsf{TINKER} will require updates as more knowledge becomes available in the form of natural language text, lest it will quickly get stale.
\subsection{Ontology}
Prior to extracting syntactic and lexical text patterns, we need to ensure that they also map to classes (entities) defined by an malware ontology. An ideal open-source ontology can systematically capture cyber threat and attack information (facts and analysis) to model the contents of threat reports. Constructing an ontology requires answering competency questions that fulfill the requirements for requisite coverage of the threat domain. These questions (or a narrower version of a competency question) can be answered by running SPARQL queries on the knowledge graph. While building \textsf{TINKER}, we updated the competency questions to the following:
\begin{enumerate}
    \item What are missing pieces of information about an attacker/ attacker-group, a malware, a malware campaign, vulnerabilities exploited or attack methods deployed.\label{CQ1}
    \item What are the similar features for grouping adversaries, malware to help understand their behavior and predict the future course of action?\label{CQ2}
    \item What is the impact of a given malware on an organization or industry -- financial, human life, intellectual property, reputation?\label{CQ3}
    \item What can we predict about the future course of action of a malware, attacker/ attacker-group, both short-term and long-term?\label{CQ4}
\end{enumerate}

\begin{figure}[htp]
        \centering
        \frame{\includegraphics[width=0.25\textwidth,height=0.28\textwidth]{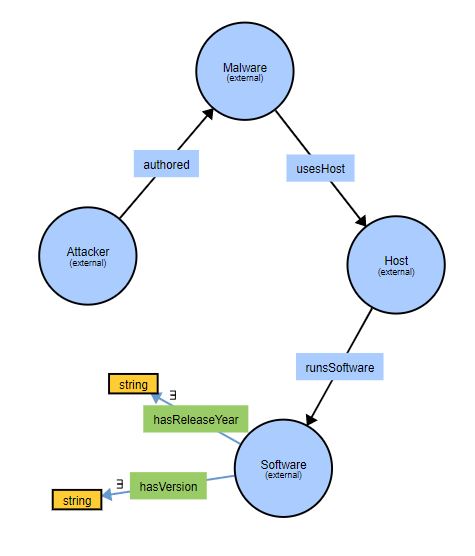}}
        \caption{Classes from a malware ontology}
        \label{fig:snap}
\end{figure}

An ontology scopes concepts within a specific domain using classes and properties [anonymous]. For the ontology to be extensible and scalable, a few classes cover the domain with a pre-defined purview. Instances hold different values for individual classes and are connected to each other by properties. Furthermore, software agents can utilize such an ontology to generate malware KGs.  Ontology classes and properties can significantly enhance past, current and future threat analysis that can inform further investigation on a variety of malware attacks. For instance, in [anonymous] ontology, three classes largely describe malware behavior -- \textsf{Malware}, \textsf{Vulnerability}, and \textsf{Indicator}. Instances of these classes can equip the analysts with considerable information on malware behavior and its attacking tactics. See Figure \ref{fig:snap} for a snapshot on a malware ontology.

\section{Generating Triples}
In this paper, we concern with manual named entity recognition (NER) and relation extraction (RE) and discuss partially automated extraction of some classes. We construct \textsf{TINKER} corresponding to the malware ontology by populating it with instances derived from actual threat reports. The knowledge graph can be queried \cite{rastogi2020malont} using competency questions written using SPARQL queries to help answer several research questions. However, one may argue about the need for knowledge graphs given that the instantiated ontology is already capturing facts and information about the domain. The ability to reason over extracted information and infer latent information are the two key features of knowledge graphs. \textsf{TINKER} captures a web of properties between individual nodes (also called entities) and uses a reasoner to draw connections between entities that would otherwise not be understood. Here, \textsf{TINKER} holds instances of specific malware attributes such as origination, attack goals, timeline, affiliated actors, vulnerabilities exploited for the attack, impact on industries as well as on humans. \textsf{TINKER} is not just an instantiation of data using an ontology. It is a web of properties between individual nodes (also called entities) and uses a reasoner to draw connections between entities that would otherwise be overlooked.

\subsection{Manual Generation}
Some of the threat reports\footnote{https://tinyurl.com/y9shcvpd} that were published by security organizations between the years 2006 to 2021 were hand-annotated using malware threat intelligence ontologies \cite{rastogi2020malont}\cite{swimmer}. For example, a 2011 report on Night Dragon\footnote{https://tinyurl.com/y5veq59m} describes it operations. Other reports published in 2013\footnote{https://tinyurl.com/y52axjtf} on malwares like Stuxnet, Shamoon allow a more in-depth and broader range of details on a particular malware attack.

\begin{figure}[htp]
        \centering
       \includegraphics[width=.28\textwidth, height=.3\textwidth]{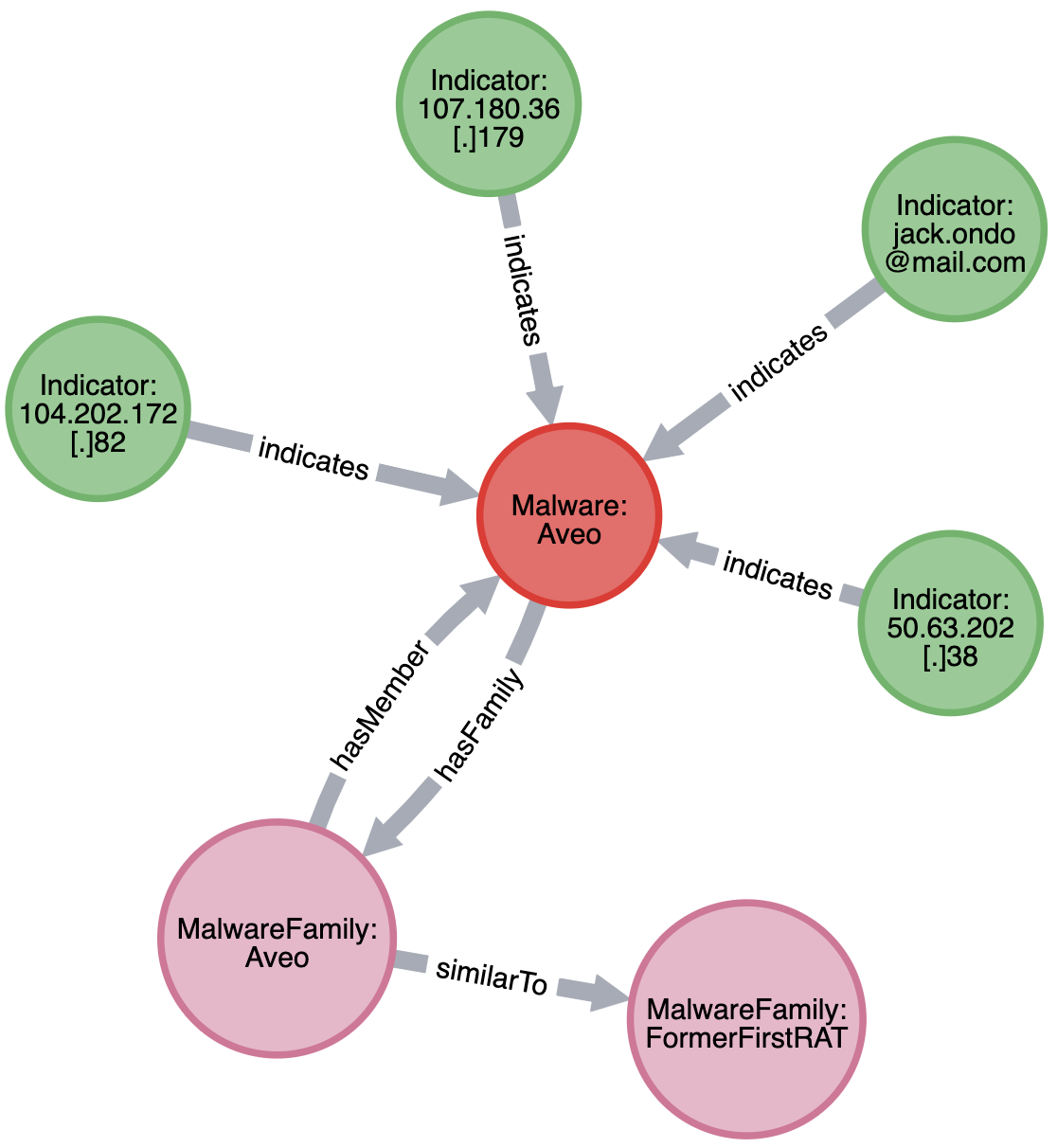}
        \caption{Snippet of \textsf{TINKER} using malware ontology. Tool used - neo4j}
        \label{fig:kg}
\end{figure}
\par
Threat reports were manually annotated by the authors and reviewed by a security expert. The annotated values from the threat reports were used to instantiate various concepts of the malware ontology. In this step, values are assigned to instances of malware ontology classes and properties. In Figure \ref{fig:kg} the snippet of the malware KG depicts modeling of threat data collected from reports using classes and properties from the malware ontology.

\subsection{Automatic Generation}
Our ultimate research goal is to develop and train our own NER and RE models. Automating the annotation process for threat reports requires an ensemble of machine learning and text mining approaches. The requirement is to extract salient key-phrases that are representative of the threat report. Therefore, we considered our research requirements and explored preexisting NER models from spaCy, Stanford, and Flair. The most common entities extracted from these NER models include - people, organization, location, date, software, product. We tested several of these models for the cybersecurity dataset, and the results had some interesting revelations. Flair12 gave us the most accurate results for the classes which were mapped back to the malware ontology.






\section{Empirical Results}
83 threat report written in unstructured textual format were converted from pdf or html to text. This formed the training and test data with words and corresponding tokens. Each report was separated into individual sentences and fed into Brat annotation tool. However, for automation, we plan to move ahead with caution and selectively annotate entities and relations. From the hand-annotated dataset, we observed that only the the entities that formed 95\% of all the annotations are capturing most of the data in the threat reports.



\section{Discussion on Ongoing Research}
To create a large scale \textsf{TINKER}, we are exploring an ensemble of information extraction techniques that have different individual strengths. The DBpedia knowledge base can annotate general-purpose key-phrases and recognize entities present in the "DBP" namespace. We incorporate existing vulnerability databases - CVE and NVD and other malware ontology along with \cite{swimmer, rastogi2020malont} to ensure we capture as much threat intelligence as possible. Besides, we used a simple regular expression to extract instances such as indicators of compromise. For annotating static information, we are relatively optimistic about Flair12. The biggest challenge lies in annotating contextual information and defining relations between classes (or entities). One of the approaches we have explored and has given reasonably good results is based on a context feature selection method. It selects clean context features for calculating entity-entity distributional similarity. However, the limitation of this approach is that it extracts only one entity of a kind from every threat report and not all instances. Another challenge that we're looking into overcoming is a document level relation extraction technique that can be trained using \textsf{TINKER}.

\bibliographystyle{acl_natbib}
\bibliography{anthology,acl2021}


\end{document}